\documentclass[iop]{emulateapj-rtx4}

\newcommand{\NH}{$N_{\rm H}$}
\newcommand{\Msun}{$M_{\odot}$}
\newcommand{\chisq}{$\chi ^2$/dof}

\shorttitle{Elemental Abundances in G344.7--0.1}
\shortauthors{Yamaguchi et al.}

\begin{document}

\title{Elemental Abundances in the Possible Type~Ia Supernova Remnant G344.7--0.1}

\author{H.\ Yamaguchi\altaffilmark{1,2}}
\email{hyamaguchi@head.cfa.harvard.edu}
\author{M.\ Tanaka\altaffilmark{3,4}}
\author{K.\ Maeda\altaffilmark{4}}
\author{P.\ O.\ Slane\altaffilmark{1}}
\author{A.\ Foster\altaffilmark{1}}
\author{R.\ K.\ Smith\altaffilmark{1}}
\author{S.\ Katsuda\altaffilmark{2}}
\author{R.\ Yoshii\altaffilmark{2}}


\altaffiltext{1}{Harvard-Smithsonian Center for Astrophysics, 60 Garden Street, 
	Cambridge, MA 02138, USA}
\altaffiltext{2}{RIKEN (The Institute of Physical and Chemical Research), 
  2-1 Hirosawa, Wako, Saitama 351-0198, Japan}
\altaffiltext{3}{National Astronomical Observatory of Japan, 
	2-21-1 Osawa, Mitaka, Tokyo 181-8588, Japan}
\altaffiltext{4}{Institute for the Physics and Mathematics of the Universe, 
  University of Tokyo, 5-1-5 Kashiwanoha, Kashiwa, Chiba 277-8568, Japan}

\begin{abstract}

Recent studies on the Galactic supernova remnant (SNR) G344.7--0.1 have 
commonly claimed its origin to be a core-collapse supernova (SN) explosion, 
based on its highly asymmetric morphology and/or proximity to a star forming region. 
In this paper, however, we present an X-ray spectroscopic study of this SNR 
using {\it Suzaku}, which is supportive of a Type~Ia origin. 
Strong K-shell emission from lowly ionized Fe has clearly been detected, 
and its origin is determined, for the first time, to be the Fe-rich SN ejecta.  
The abundance pattern is highly consistent with that expected for a 
somewhat-evolved Type Ia SNR. It is suggested, therefore, that the X-ray point-like source 
CXOU~J170357.8--414302 located at the SNR's geometrical center is not associated 
with the SNR but is likely to be a foreground object. 
Our result further indicates that G344.7--0.1 is the first possible Type~Ia SNR 
categorized as a member of the so-called ``mixed-morphology" class. 
In addition, we have detected emission from He-like Al at $\sim$1.6~keV, 
the first clear detection of this element in the spectrum of an extended X-ray source. 
The possible enhancement of the Al/Mg abundance ratio from the solar value 
suggests that the ambient interstellar medium has a relatively high metallicity 
(not less than 10\% of the solar value).  
We also report the marginal detection of Cr and Mn, although the measured fluxes 
of these lines have large statistical and systematic uncertainties. 
 
\end{abstract}


\section{Introduction}
\label{sec:introduction}

\begin{figure*}[t]
  \begin{center}
      \includegraphics[scale=.26]{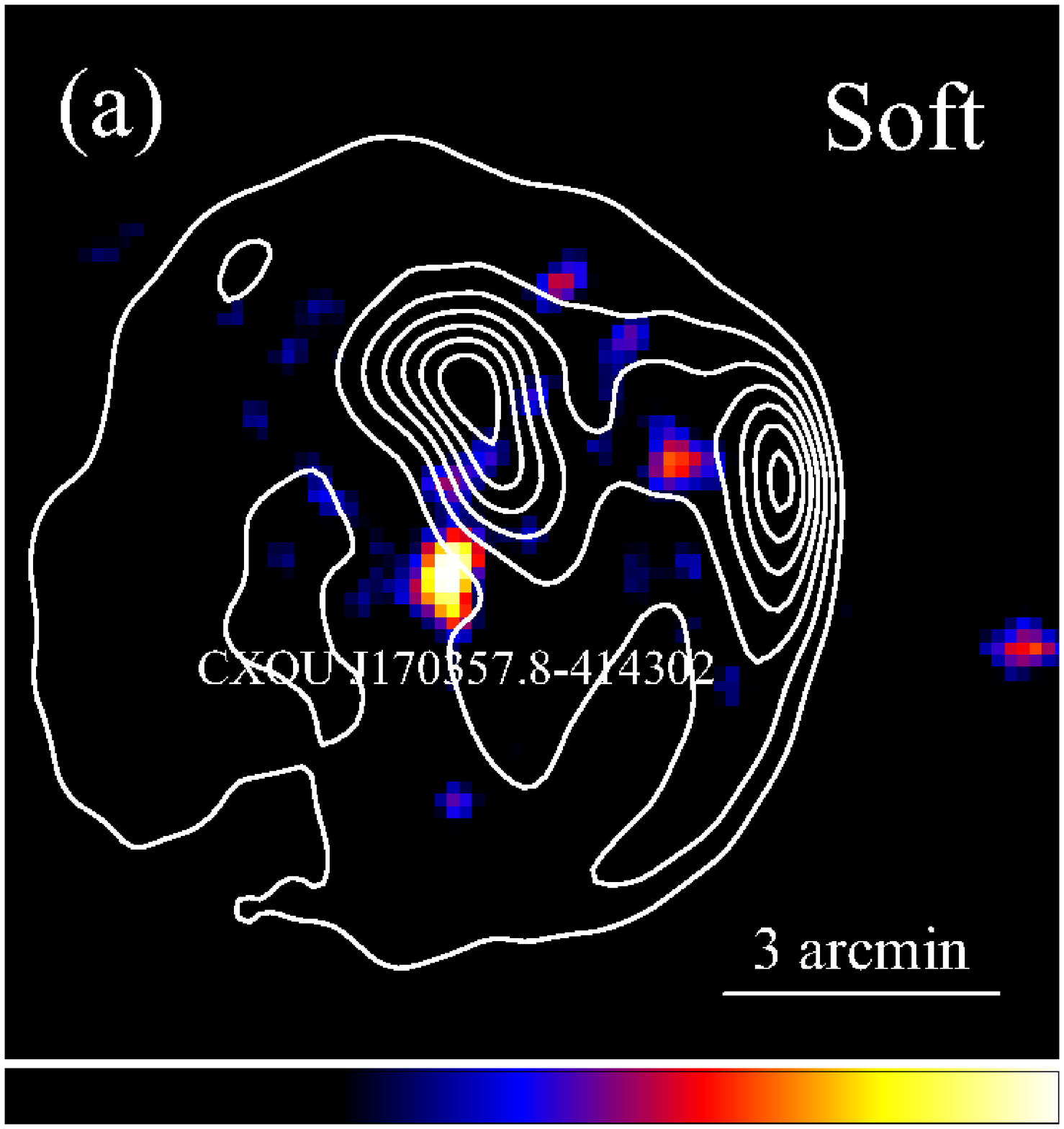}
      \includegraphics[scale=.26]{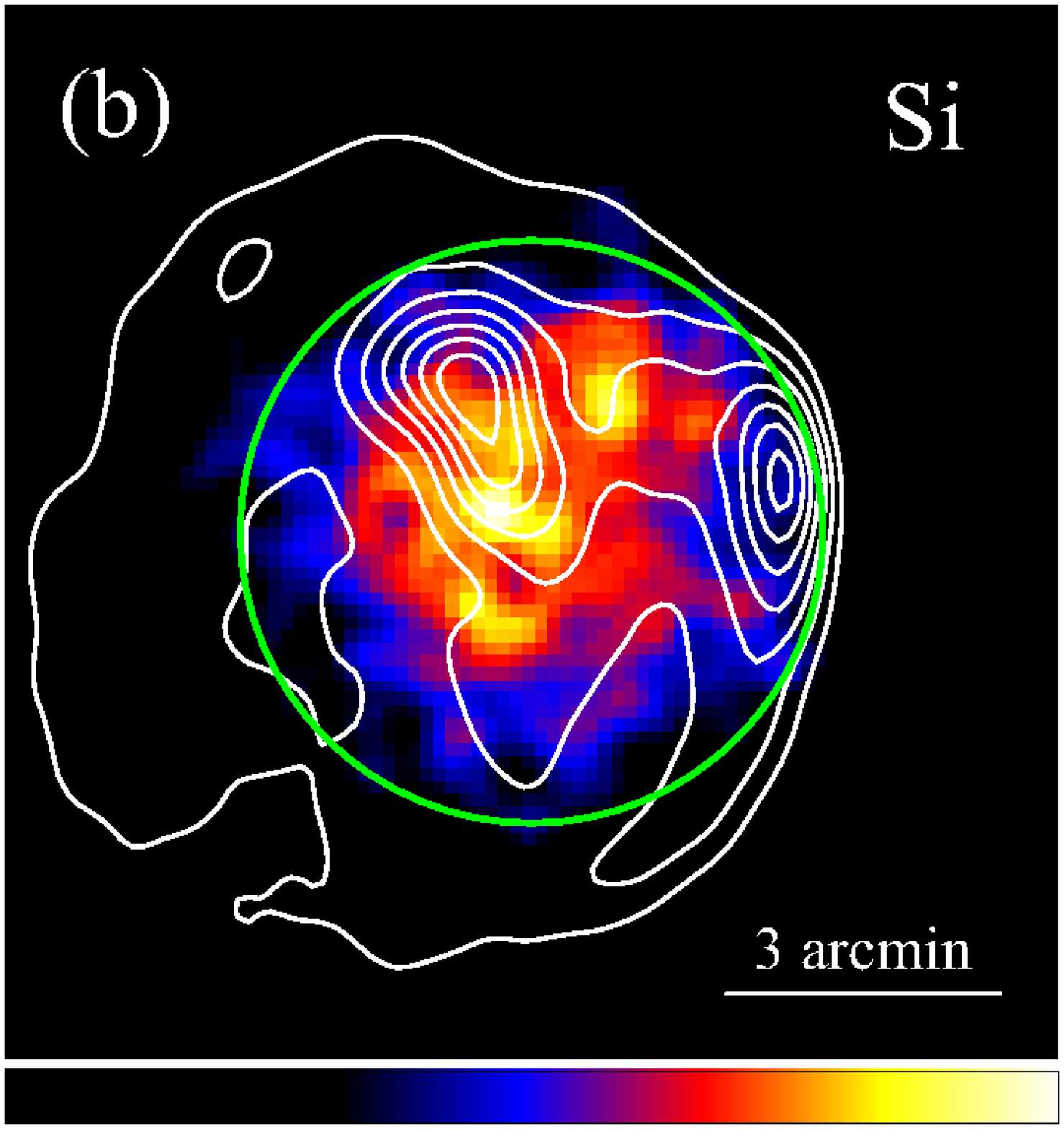}
      \includegraphics[scale=.26]{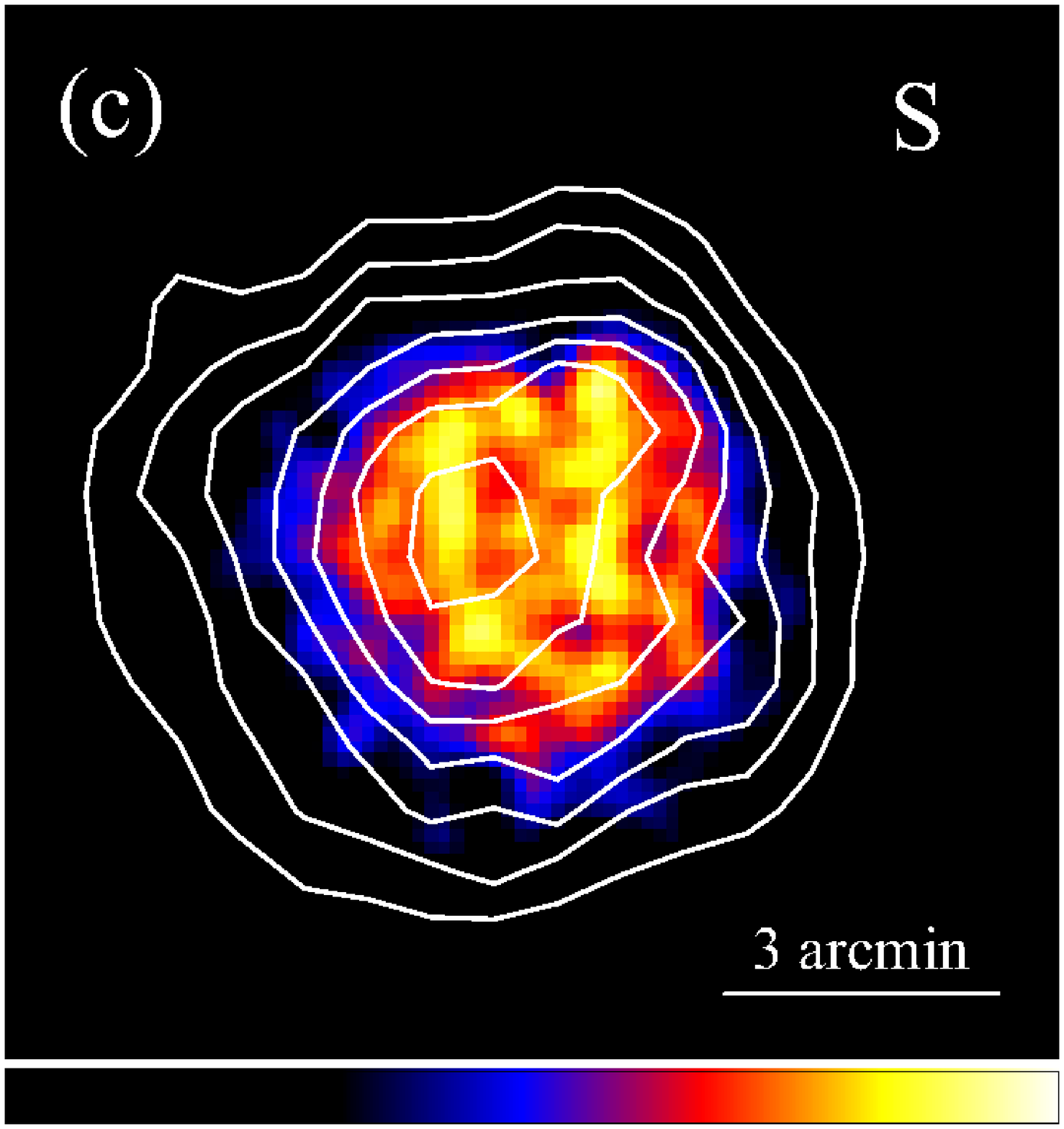}\\
      \includegraphics[scale=.26]{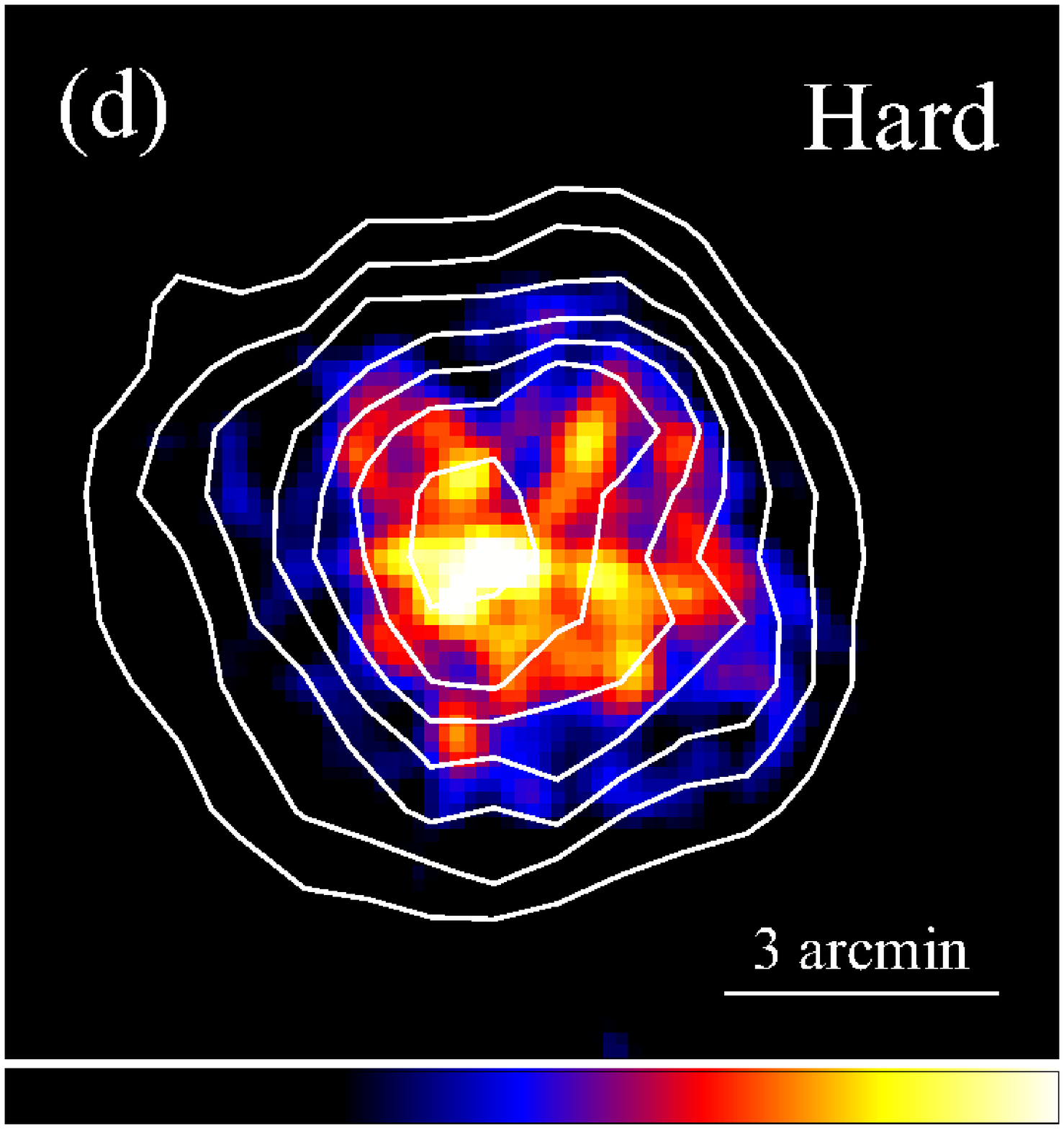}
      \includegraphics[scale=.26]{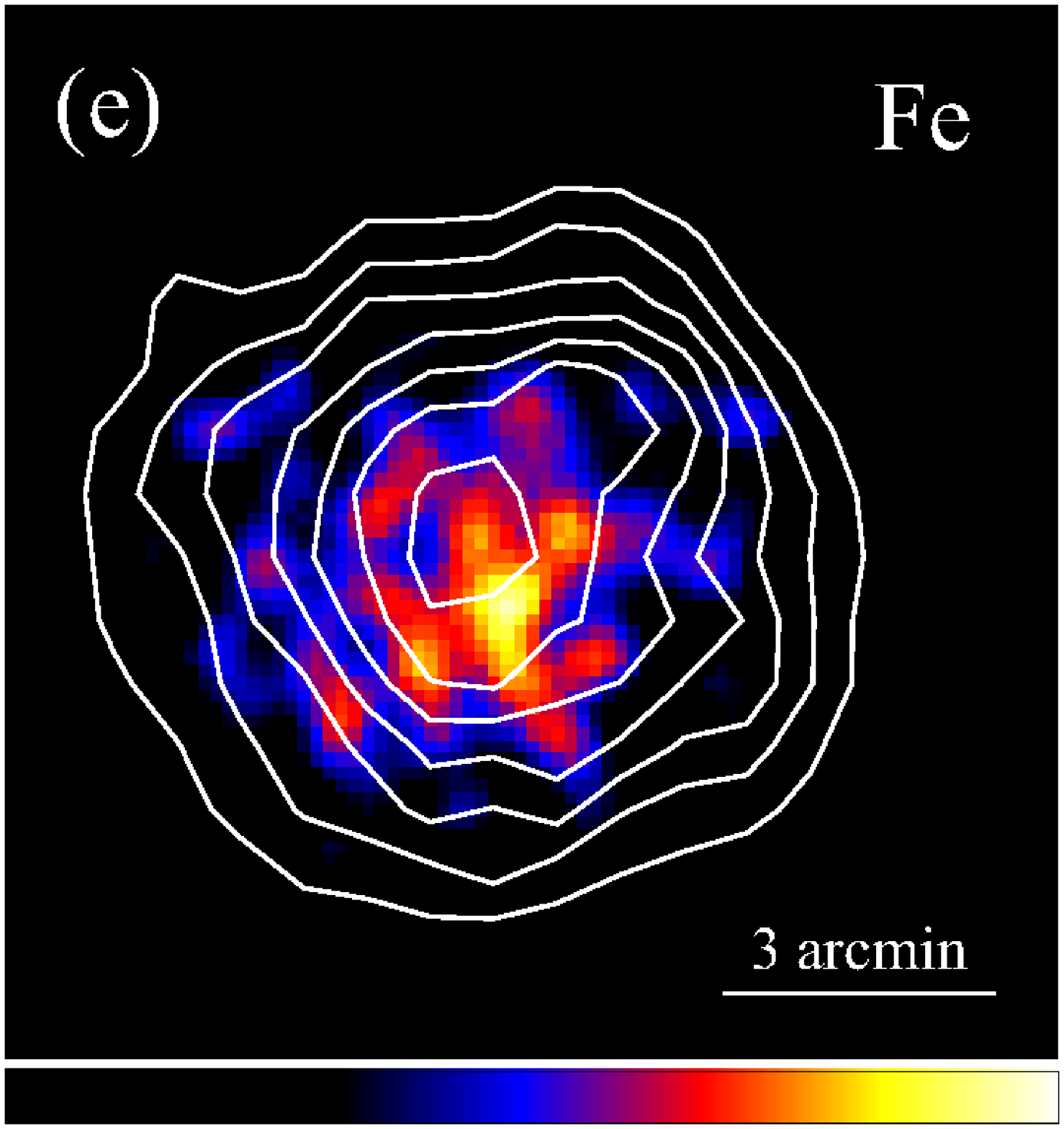}
      \includegraphics[scale=.26]{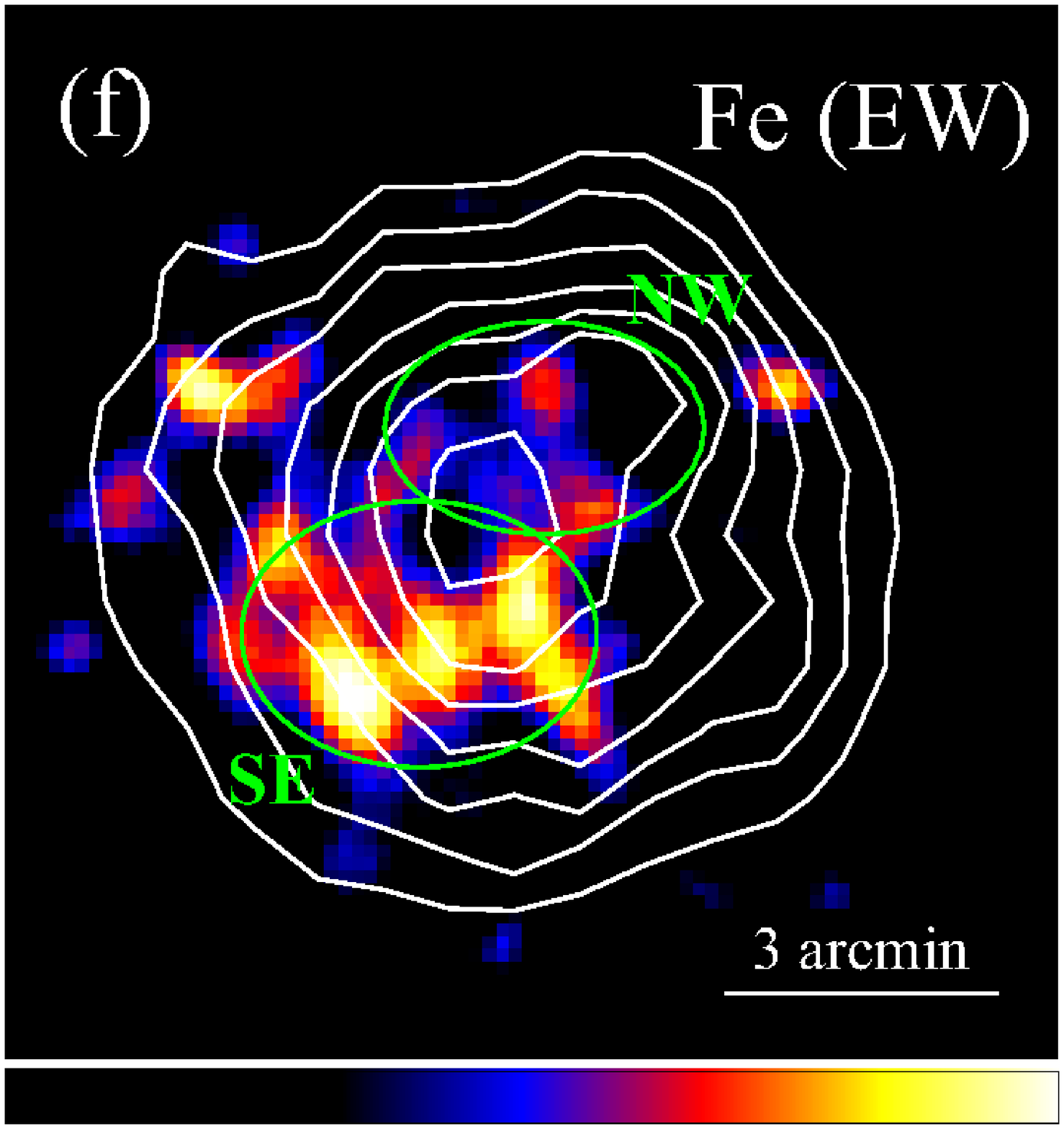}
\caption{XIS images of G344.7--0.1 in the energy bands of 
	the soft continuum (a: 0.7--1.2~keV), Si-K (b: 1.81--1.90~keV), 
	S-K (c: 2.41--2.51~keV), hard continuum (d: 4.2--6.0~keV), and 
	Fe-K (e: 6.34--6.57~keV), and the equivalent-width map of 
	the Fe-K emission (f). North is up and east is to the left. The scale is linear. 
	Intensity contours for the {\it MOST} 843~MHz radio emission 
	(Whiteoak \& Green 1996) are shown in the images (a) and (b), 
	whereas those for the Si-K emission are overlaid on the images (c), (d), (e), and (f). 
	The regions used in the spectral analysis are indicated in (b); 
	the green circle is centered on the position (R.A., decl.)$_{\rm J2000}$ = 
	(17$^{\rm h}$03$^{\rm m}$50.7$^{\rm s}$, --41$^{\circ}$42$'$41$''$). 
  \label{fig:image}}
  \end{center}
\end{figure*}

The supernova remnant (SNR) G344.7--0.1 was discovered in radio observations  
(Caswell et al.\ 1975) and exhibits largely asymmetric structure with a diameter of 
$\sim$$10'$ and a bright western part (Dubner et al.\ 1993; Whiteoak \& Green 1996). 
In X-rays, the SNR was first identified by the {\it ASCA} Galactic Plane Survey 
(Sugizaki et al.\ 2001). \cite{y05} found that the {\it ASCA} spectrum can be 
reproduced by an optically-thin thermal plasma with an electron temperature of 
$\sim$1~keV. Owing to the limited photon statistics, however, they could not determine 
whether the plasma had achieved collisional ionization equilibrium (CIE) or was still in 
non-equilibrium ionization (NEI). Another interesting result from {\it ASCA} was the possible 
detection of Fe-K$\alpha$ emission with an energy of 6.4 ($\pm$ 0.2)~keV. 
Its origin was suggested to be either fluorescence from cold dense clouds
or low-ionized plasma, but was not definitively determined.

The X-ray morphology of the SNR was examined in detail during {\it Chandra} 
and {\it XMM-Newton} observations. The emission correlates relatively well with 
the radio structure \citep{c10}, but the X-ray peak is located more 
interior with respect to that in the radio \citep{g11}. It was argued 
that the irregular morphology of this SNR may be due to expansion through 
a dense interstellar medium (ISM) with a density gradient toward the western part.  
\cite{g11} suggested that this dense environment interacting with the SNR's shock 
is a plausible counterpart of the unidentified TeV $\gamma$-ray source 
HESS J1702--420 (Aharonian et al.\ 2008). 
Notably, \cite{c10} discovered a point-like source, CXOU J170357.8--414302, 
at the geometrical center of the SNR with no significant extended emission 
to indicate a pulsar wind nebula. The source was, therefore, suggested to be 
a possible candidate for a compact central object (CCO) physically associated 
with the SNR. However, the absorption column of the point source was found to be 
significantly smaller than that of the SNR, implying that the source is a foreground 
object. In fact, the infrared/optical object coinciding with CXOU J170357.8--414302 
exhibits typical characteristics of an early-K star \citep{c10}.

The progenitor of G344.7--0.1 is currently not well identified, but most of the previous 
works are supportive of a core-collapse (CC) origin, based mainly on its associations 
with a nearby molecular cloud or a wind-blown bubble 
(Combi et al.\ 2010; Giacani et al.\ 2011). 
In fact, infrared emission from shocked molecular gas has been observed in 
this SNR, which is more common in CC SNRs than in Type~Ia remnants 
(Reach et al.\ 2006; Andersen et al.\ 2011). 
Furthermore, it has recently been argued that its highly asymmetric X-ray morphology 
is somewhat similar to other CC SNRs \citep{l11}. 
In these studies, however, the abundances of heavy elements could not be measured 
accurately, and the detection of the Fe K-shell emission was not clear or only marginal. 
Since production of Fe ($^{56}$Ni, before nuclear decay) is significantly different 
between CC and Type~Ia supernovae (SNe) (e.g., Iwamoto et al.\ 1999), clear detection 
and diagnostics of the Fe emission as well as accurate measurements of the elemental 
abundances could help resolve the nature of its progenitor and environment.

In this paper, we present the results of the {\it Suzaku} observation of G344.7--0.1, 
utilizing its high spectral sensitivity especially in the Fe K-shell X-ray (6--7~keV) band. 
The errors quoted in the text and tables are at the 90\% confidence level, 
and the error bars given in the spectra are for 1$\sigma$ confidence, 
unless otherwise stated.

\section{Observation and Results}
\label{sec:result}


G344.7--0.1 was observed by {\it Suzaku} on 2007 February 21, 
using X-ray Imaging Spectrometers (XISs). 
Each XIS consists of three active X-ray sensors. 
Two of them are installed with front-illuminated (FI) sensors, while the other is 
installed with a back-illuminated (BI) sensor. 
We used the standard tools of \texttt{HEADAS} version 6.11 for the data reduction. 
The archival data were reprocessed with the calibration database released on 
2011 August 11 and reduced in accordance with the recommended screening 
criteria.$\!$\footnote{http://heasarc.nasa.gov/docs/suzaku/processing/criteria\_xis.html} 
The total effective exposure time after the screening was about 44~ks. 
Only grade 0, 2, 3, 4, and 6 events were used in the following analysis.

\subsection{Imaging Analysis}
\label{ssec:image}


In Figure~\ref{fig:image}, we present narrow-band images of the SNR 
in the energy ranges of (a) 0.7--1.2~keV, (b) 1.81--1.90~keV, (c) 2.41--2.51~keV, 
(d) 4.2--6.0~keV, and (e) 6.34--6.57~keV (hereafter, ``soft", ``Si-K", ``S-K", 
``hard", and ``Fe-K", respectively). 
In the soft image, we can see a point-like source located nearly at the center 
of the remnant. The peak position of the source is (R.A., decl.)$_{\rm J2000.0}$ = 
(17$^{\rm h}$03$^{\rm m}$56$^{\rm s}$, --41$^{\circ}$42$'$56$''$), 
consistent with the location of CXOU J170357.8--414302 within the 
position uncertainty of the XIS ($\sim$$1'$; Uchiyama et al.\ 2008).

To compare the surface brightness distribution among the different energy bands, 
intensity contours for the Si-K emission are superimposed on the images in 
Figure~\ref{fig:image}(c), (d), (e), and (f). 
The spatial profiles of the S-K and hard continuum emission are highly 
correlated with the Si-K image, whereas the peak of the Fe-K emission 
is slightly shifted toward the south.  
An equivalent-width (EW) map for the Fe-K line is shown in Figure~\ref{fig:image}(f), 
where the continuum level was estimated by scaling the hard image with 
a bremsstrahlung model fitting for the 4.2--6.0~keV spectrum. 
The Fe-K EW is enhanced at the southeastern part of the remnant, 
and not spatially correlated with the Si-K emission.

\subsection{Spectral Analysis}
\label{ssec:spectrum}

\begin{figure}[t]
  \begin{center}
      \includegraphics[scale=.50]{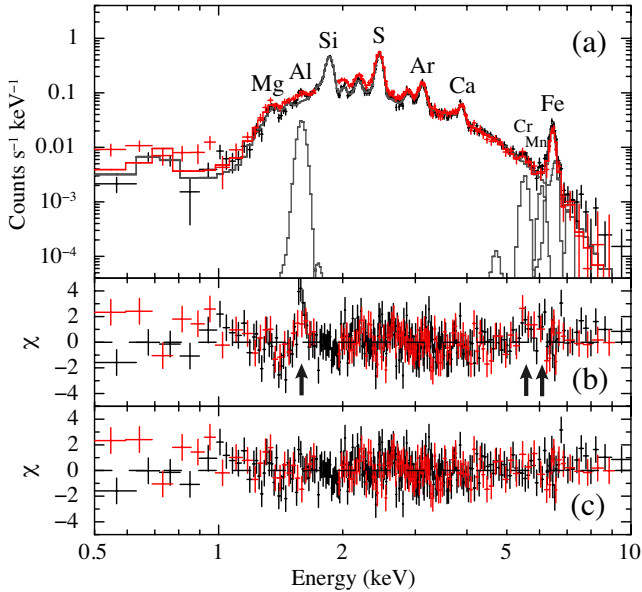}
\caption{(a) XIS spectra of the entire SNR.  
	Black and red represent the FI and BI spectra, respectively. 
	Individual components of the best-fit model are shown with gray lines. 
	(b) and (c) display the residuals from the models {\it without} and {\it with} 
	the Gaussian components of Al-K$\alpha$, Cr-K$\alpha$, and Mn-K$\alpha$, respectively. 
	Note that the energies below 1.0~keV are ignored from the spectral fitting. 
  \label{fig:full}}
  \end{center}
\end{figure}

We extracted the XIS spectra of the entire SNR from a circular region with 
a radius of 3.$\!'$2 (shown in Figure~\ref{fig:image}(b)), 
but the central point source was excluded. 
The two FI spectra were summed up to improve the photon statistics, 
since their characteristics are almost identical. 
The background data were taken from an annulus surrounding the SNR 
with inner and outer radii of 5.$\!'$0 and 7.$\!'$0, respectively.
The resultant spectra are shown in Figure~\ref{fig:full}(a). 
We see the Fe-K$\alpha$ line detected with good statistics. 
Using {\tt XSPEC} (Arnaud 1996), we first fitted the spectra with an absorbed 
NEI plasma model, {\tt wabs} {\tt $\times$} {\tt vnei} (K.\ Borkowski\footnote{This 
model is based on the ion balance and emissivity calculations by Astrophysical 
Plasma Emission Code (APEC; Smith et al.\ 2001), but emission caused by 
innershell ionization and excitation is included as well. Detailed information is 
found at http://space.mit.edu/home/dd/Borkowski/APEC\_nei\_README.txt.}). 
The free parameters were the absorption column density (\NH), electron temperature 
($kT_e$), ionization timescale ($n_et$), volume emission measure (VEM), 
and elemental abundances of Mg, Si, S, Ar, Ca, and Fe relative to the solar 
values of Anders \& Grevesse (1989). The photo-electric absorption cross sections 
were taken from Morrison \& McCammon (1983). 
The gain calibration of the XIS is known to be problematic around the energy of 
the neutral Si K-edge (e.g., Yamaguchi et al.\ 2009). Therefore, we ignore the energy 
range of 1.75--1.90~keV in the BI spectrum, similar to Yamaguchi et al.\ (2009). 
In addition, a significant inconsistency between the FI and BI data was found 
below 1~keV. Although we also attempted to use background data from a blank sky 
region and pure non X-ray background data, instead of the data taken from the outer 
annulus region, the inconsistency remained in all cases. 
We conclude that the problem is not caused by the background subtraction but is 
due to issues related to the source spectra. 
One of the possible origins is an incomplete calibration of the thickness of contamination 
material accumulated on the optical blocking filters of the XIS (Koyama et al.\ 2007). 
Although this is not definitive, we excluded the data below 1.0~keV from 
the following analysis.

The model described above gave a good fit to the emission lines at energies 
below 4~keV and the overall continuum spectrum, but failed to reproduce 
the Fe-K$\alpha$ line profile; the peak energy was lower in the data than 
in the model. Although we temporarily used a plane-parallel shock model 
({\tt vpshock}; Borkowski et al.\ 2001) instead of the simple {\tt vnei} model, 
no significant improvement was achieved. 
This implies that the residual Fe-K$\alpha$ emission arises from either another 
plasma component with a lower ionization timescale or fluorescence of neutral matter. 
To reveal its nature, we examined the Fe-K$\alpha$ line parameters with an additional 
Gaussian model. Hereafter, the Fe abundance in the {\tt vnei} component was 
fixed to unity, as for the other fixed parameters (C, N, O, and Ne). 
The center energy and EW of the additional Fe-K$\alpha$ component 
were obtained to be 6.44 ($\pm 0.01$)~keV and 3.7 ($\pm 0.4$)~keV, respectively. 
Since the former value is significantly different from the energy of K$\alpha$ emission 
from neutral Fe (6.40~keV), we can safely reject the fluorescence scenario. 
Furthermore, the EW corresponds to an Fe abundance higher than 10~solar 
for any electron temperature, implying that the predominant origin of the emission is 
likely to be the ejecta rather than the ISM. Interestingly, the measured Fe-K$\alpha$ 
centroid is fully consistent with that observed in Tycho's SNR (Tamagawa et al.\ 2009), 
suggesting a similar ionization state of Fe in both SNRs. 
Tamagawa et al.\ (2009) also reported the detection of the Fe-K$\beta$ line in the 
{\it Suzaku} spectrum of Tycho's SNR, with K$\beta$/K$\alpha$ intensity ratio of 
$\sim$0.05. We followed this result to add another Gaussian at 7.11~keV by assuming 
the same intensity ratio. The best fit thereby obtained had $kT_e$ = 1.0~keV 
and \chisq\ = 410/309. 
The residual with this model is shown in Figure~\ref{fig:full}(b).

Although the fit is now almost acceptable, a prominent line-like residual is 
still observed at $\sim$1.6~keV (in both the FI and BI spectra). 
We tried two-component {\tt vnei} models, but this residual did not disappear even 
with independent free parameters including abundances between the components. 
Given that the {\tt vnei} component already takes into account the K$\beta$ emission 
of Mg XI (1.58~keV), the most likely origin of this feature is Al XII K$\alpha$ blend 
(e.g., 1.60~keV for the resonance line) 
that is not included in the {\tt vnei} code in {\tt XSPEC}. 
Introducing a Gaussian to account for this emission, we achieved 
a significant reduction in the \chisq\ value (from 410/309 to 354/306). 
In addition, we still see small residuals at $\sim$5.5~keV and $\sim$6.1~keV 
(indicated by black arrows in Figure~\ref{fig:full}(b)), 
possibly due to Cr- and Mn-K$\alpha$ emissions, respectively. 
Adding a Gaussian for the former resulted the reduction in $\chi ^2$ of 15, 
while that for the latter resulted the additional reduction of 9. 
The best-fit parameters and residuals from the model obtained after adding these 
emission features are given in the ``Model~Ax" column in Table~\ref{tab:best} and 
Figure~\ref{fig:full}(c), respectively.

\begin{table*}[t]
\begin{center}
\caption{Spectral parameters.
  \label{tab:best}}
\begin{tabular}{llcccc}
  \tableline\tableline
 Component  & Parameter   &  Model~Ax  &  Model~Bx  &  Model~As  &  Model~Bs \\
  \tableline
  Absorption$^{a}$ & \NH ($10^{22}$ cm$^{-2}$) & 4.7$_{-0.2}^{+0.2}$ & 5.1$_{-0.3}^{+0.3}$ & 4.6$_{-0.1}^{+0.1}$ & 4.6$_{-0.1}^{+0.2}$ \\
  NEI plasma$^{b}$  &  $kT_e$ (keV)  &  0.98$_{-0.03}^{+0.04}$ & 0.92$_{-0.04}^{+0.04}$ & 1.01$_{-0.04}^{+0.04}$ & 0.96$_{-0.06}^{+0.05}$ \\
  ~ & Mg (solar) & 0.52$_{-0.15}^{+0.18}$ & 0.87$_{-0.30}^{+0.37}$ & 0.66$_{-0.15}^{+0.18}$ & 0.60$_{-0.14}^{+0.17}$  \\
  ~ & Al (solar) & --- & --- & 1.9$_{-0.4}^{+0.5}$ & 1.6$_{-0.4}^{+0.4}$  \\
  ~ & Si (solar) & 1.1$_{-0.1}^{+0.1}$ & 1.3$_{-0.2}^{+0.2}$ & 1.3$_{-0.1}^{+0.1}$ & 1.1$_{-0.1}^{+0.1}$ \\
  ~ & S (solar)  & 1.9$_{-0.1}^{+0.1}$ & 2.0$_{-0.1}^{+0.1}$ & 2.2$_{-0.1}^{+0.1}$ & 2.0$_{-0.1}^{+0.1}$ \\
  ~ & Ar (solar) & 2.0$_{-0.2}^{+0.2}$ & 2.1$_{-0.2}^{+0.2}$ & 2.3$_{-0.2}^{+0.2}$ & 2.2$_{-0.2}^{+0.2}$ \\
  ~ & Ca (solar) & 2.2$_{-0.2}^{+0.2}$ & 2.2$_{-0.3}^{+0.3}$ & 2.8$_{-0.4}^{+0.4}$ & 2.7$_{-0.4}^{+0.4}$ \\  
  ~ & $n_et$ ($10^{11}$ cm$^{-3}$ s) & 1.7$_{-0.2}^{+0.2}$ &  2.2$_{-0.3}^{+0.5}$ & 1.2$_{-0.2}^{+0.2}$ & 1.3$_{-0.2}^{+0.2}$ \\
  ~ & VEM ($10^{12}$ cm$^{-5}$)$^{c}$  & 7.0$_{-0.8}^{+0.7}$ & 8.0$_{-0.9}^{+0.9}$ & 6.1$_{-0.5}^{+0.7}$ & 7.2$_{-0.8}^{+1.0}$   \\
  Fe ejecta  &  $E$ (eV)  &  6447$_{-12}^{+11}$ & --- & 6455$_{-11}^{+11}$ & --- \\
  ~ & $kT_e$ (keV)  &  --- & 3.6$_{-0.8}^{+1.2}$ & --- & 2.8$_{-0.7}^{+7.0}$ \\
  ~ & $n_et$ ($10^9$ cm$^{-3}$ s) & --- & 9.3$_{-2.0}^{+2.2}$ & --- & 7.2$_{-1.1}^{+1.5}$ \\
  ~ & $\sigma$ (eV)  & 78$_{-20}^{+18}$ & 67$_{-23}^{+18}$ & 85$_{-22}^{+20}$ & 75$_{-19}^{+17}$ \\
  ~ & Normalization$^{d}$ & 2.8$_{-0.3}^{+0.3}$ & 1.4$_{-0.6}^{+1.2}$ & 3.0$_{-0.3}^{+0.3}$ & 2.2$_{-1.5}^{+3.5}$ \\  
  Al-K$\alpha$  & $E$ (eV) & 1585$_{-9}^{+10}$ & 1585$_{-10}^{+9}$ & --- & --- \\
 ~ & $\sigma$ (eV)  & 8$_{-8}^{+21}$ & 12$_{-12}^{+24}$ & --- & --- \\
 ~ & Flux ($10^{-4}$ ph cm$^{-2}$ s$^{-1}$) & 3.7$_{-1.0}^{+1.2}$ & 5.6$_{-1.7}^{+2.5}$ & --- & --- \\
  Cr-K$\alpha$$^{e}$  & $E$ (eV) & 5526$_{-66}^{+70}$ & 5524$_{-69}^{+72}$ & 5530$_{-68}^{+76}$ & 5527$_{-71}^{+80}$ \\
 ~ & Flux ($10^{-6}$ ph cm$^{-2}$ s$^{-1}$) & 3.6$_{-1.4}^{+1.6}$ & 3.4$_{-1.5}^{+1.5}$ & 3.6$_{-1.5}^{+1.7}$ & 3.3$_{-1.5}^{+1.5}$ \\
  Mn-K$\alpha$$^{e}$  & $E$ (eV) & 6085$_{-120}^{+107}$ & 6106$_{-126}^{+117}$ & 6076$_{-126}^{+113}$ & 6088$_{-131}^{+121}$ \\
 ~ & Flux ($10^{-6}$ ph cm$^{-2}$ s$^{-1}$) & 2.4$_{-1.3}^{+1.2}$ & 2.4$_{-1.4}^{+1.7}$ & 2.3$_{-1.4}^{+1.4}$ & 2.2$_{-1.4}^{+1.5}$ \\
  \tableline
  \multicolumn{2}{l}{$\chi ^2$/d.o.f.}  &  330/302 & 319/301 & 370/304 & 362/303 \\
  \tableline
\end{tabular}
\tablecomments{
  $^{a}$The absorption cross sections were taken from Morrison \& McCammon (1983).
  $^{b}$Solar abundances of Anders \& Grevesse (1989) are assumed. 
  The values for C, N, O, Ne, Fe, and Ni are fixed to unity. 
  $^{c}$Volume emission measure, $\int n_e n_{\rm H} dV/(4\pi D^2)$, where 
  $V$ and $D$ are the emitting volume (cm$^3$) and distance to the source (cm), respectively. 
  $^{d}$Flux ($10^{-5}$ ph cm$^{-2}$ s$^{-1}$) for Models~Ax/As and  
  $\int n_e n_{\rm Fe}dV/(4\pi D^2)$ ($10^8$ cm$^{-5}$) for Models~Bx/Bs. 
  $^{e}$Gaussian widths are linked to that of the Fe-K$\alpha$ line. 
}
\end{center}
\end{table*}

\begin{figure}[t]
  \begin{center}
      \includegraphics[scale=.50]{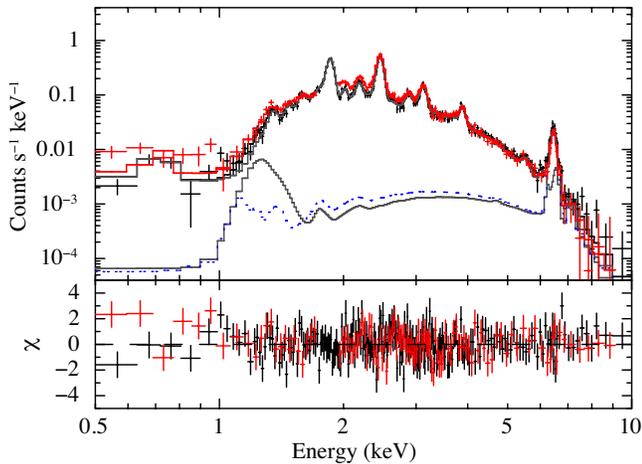}
\caption{Same as Figure~\ref{fig:full}, but a pure Fe plasma model 
	is used instead of the Fe-K$\alpha$ Gaussian. 
	The model components with solid lines and residuals are for Model~Bx.
	The blue dotted line indicates the contribution of the Fe ejecta component in 
	the best fit of Model~Bs. The Fe L-shell flux is weaker than that in Model~Bx. 
  \label{fig:full2}}
  \end{center}
\end{figure}

As a final step, we replaced the 6.44-keV Gaussian (Fe-K$\alpha$) with a {\tt vnei} 
model which consists purely of Fe ions and electrons, because the Fe-rich ejecta 
may contribute to the L-shell emission as well. We found that the Fe-K$\alpha$ 
line is broadened to a width greater than that expected for a single {\tt vnei} or 
{\tt vpshock} model, and thus introduced a {\tt gsmooth} model to convolve 
the second {\tt vnei} (Fe ejecta) component with a Gaussian with a finite width. 
In addition, since the {\tt vnei} model does not include the Fe-K$\beta$ line data, 
we left the 7.11-keV Gaussian by fixing its flux to the best-fit value from Model~Ax. 
The resultant spectral model is shown in Figure~\ref{fig:full2}, and the best-fit 
parameters are given in the ``Model~Bx" column in Table~\ref{tab:best}. 
All the values are consistent with those in Model~Ax within the statistical errors. 
This implies that neither uncertainty in Fe L-shell emissivity nor an electron temperature 
of the Fe-ejecta component crucially affects the abundance measurement.

We further analyzed the spectrum with the same procedure but using the {\tt SPEX} 
software package (Kaastra et al.\ 1996), so as to estimate the systematic 
uncertainties in the results as carefully as possible. 
Since the NEI model in {\tt SPEX} includes the Al K-shell emission, the Gaussian 
at 1.6~keV was removed. The results are given in the columns of ``Model~As" and 
``Model~Bs" in Table~\ref{tab:best}. Again, no significant variation is found among 
the results with different models or atomic codes. 
The contribution of the Fe ejecta component in Model~Bs is shown as a blue dotted 
line in Figure~\ref{fig:full2}. 
We found that the emissivity ratio of the Fe L-shell to K-shell emissions is generally 
given to be lower in {\tt SPEX} for an identical electron temperature. 
This makes limits of the temperature and normalization more difficult to determine 
in Model~Bs, since the temperature of the Fe-ejecta component is mainly determined 
by the Fe L-shell-to-K-shell intensity ratio.

The elemental abundances of the first {\tt vnei} component 
(``NEI plasma" in Table~\ref{tab:best}) are moderately higher than 
the solar values, with the exception of Mg, suggesting the origin of 
this component to be a mixture of the ejecta and swept-up ISM. 
This interpretation will be discussed more quantitatively in Section~\ref{ssec:fe}.

\section{Discussion and Concluding Remarks}
\label{sec:discussion}

\subsection{Distance to the SNR}
\label{ssec:distance}

The distance to G344.7--0.1 is highly uncertain, despite its importance for 
determination of the SNR's physical properties. 
It had once been estimated to be $\sim$14~kpc from the $\Sigma$--D relation 
(Dubner et al.\ 1993), but this phenomenological method is notoriously unreliable 
with large intrinsic dispersion. 
Based on HI absorption and emission features possibly associated with the SNR, 
\cite{g11} recently estimated a distance of 6.3 ($\pm 0.1$)~kpc. 
However, this estimate is also uncertain (see Hayato et al.\ 2010 and references 
therein for the case of the Tycho's SNR), and is suspected to be unreasonably 
small due to the following reasons.

We have obtained a foreground absorption column density of 
$N_{\rm H}$ = (4.5--5.4) $\times \, 10^{22}$~cm$^{-2}$ (Table~\ref{tab:best}), 
which is consistent with all previous results 
(Yamauchi et al.\ 2005; Combi et al.\ 2010; Giacani et al.\ 2011). 
According to the DL map (Dickey \& Lockman 1990) and the LAB map 
(Kalberla et al.\ 2005), the {\it total} Galactic HI column density $N$(HI) toward 
the direction of G344.7--0.1 is given to be $1.8 \times 10^{22}$~cm$^{-2}$ 
and $1.5 \times 10^{22}$~cm$^{-2}$, respectively. 
It is known that averaged number densities of molecular hydrogen and neutral 
atomic hydrogen are comparable with each other in the Galactic plane, 
while that of ionized hydrogen is negligibly small (e.g., Ferri{\`e}re 2001). 
We hence obtain a total Galactic hydrogen column density of $N$(total) 
= $N$(HI) + 2$N$(H$_2$) = (4.5--5.4) $\times \, 10^{22}$~cm$^{-2}$. 
This value is fully consistent with the measured absorption column for 
G344.7--0.1, suggesting that the SNR is situated near the opposite edge 
of the Galactic plane, or at least farther than the Galactic tangent point 
of this direction (8.2~kpc; a distance to the Galactic center is assumed 
to be 8.5~kpc). In fact, the distances to SNRs CTB~37A (G348.5+0.1) and 
CTB~37B (G348.7+0.3), located in a similar direction, are respectively known 
to be $\sim$11~kpc (Reynoso \& Mangum 2000) and $\sim$10~kpc 
(Caswell et al.\ 1975), but have smaller $N_{\rm H}$ values of 
(3--4) $\times \, 10^{22}$~cm$^{-2}$ (Sezer et al.\ 2011; Nakamura et al.\ 2009). 
Although \cite{g11} also performed similar estimates to support their claim, 
they assumed that 2/3 of the total Galactic hydrogen (and associated heavy 
elements) is distributed in the SNR's foreground. This assumption is quite 
unreasonable, given that the claimed distance of 6.3~kpc is closer than 
the Galactic tangent point. 
Therefore, the distance to the SNR is assumed to be 14~kpc 
in the following discussion, even though this value is not definitive either.

\subsection{Fe Emission and Possible Progenitor}
\label{ssec:fe}

We have clearly detected strong Fe K-shell emission and determined its origin 
to be the SN ejecta, for the first time. The Fe ejecta can be characterized by 
a pure-metal plasma whose electron temperature and ionization timescale are 
different from those of the other plasma component with abundant 
$\alpha$-elements (i.e., Mg, Si, S, Ar, and Ca). 
\cite{c10} argued that the wideband X-ray spectrum, including the Fe-K$\alpha$ emission, 
can be well represented by a one-component \texttt{vpshock} model 
with $kT_e$ = 1.17~keV and $\tau_u$ (a maximum value of ionization timescales) = 
$2.5\times 10^{11}$~cm$^{-3}$~s. However, photon statistics of the {\it Chandra} 
and {\it XMM-Newton} data are too low around the Fe-K$\alpha$ emission to 
independently determine its center energy. We found that the \texttt{vpshock} model 
applied by \cite{c10} predicts an Fe-K$\alpha$ centroid of $\sim$6.57~keV. 
This value is significantly higher than the energy derived from the {\it Suzaku} spectra, 
and hence at least two components are absolutely needed. 

In addition to the plasma condition, the emitting volumes of both components 
are also likely to be different from each other (see Figure~\ref{fig:image}).
Therefore, direct comparison of the abundances (and hence masses) among 
the Fe and $\alpha$-elements is not straightforward, although this comparison 
is the most credible way to distinguish the SN type. 
However, we should note that strong Fe emission is observationally very common 
in Type~Ia SNRs and is unusual in CC SNRs (e.g., Hughes et al.\ 1995). 
Furthermore, the ionization timescale for the Fe ejecta in G344.7--0.1 is found to be about 
one order of magnitude lower than that for the other elements (Table~\ref{tab:best}), 
which is another typical characteristic of Type~Ia SNRs (Hwang et al.\ 1998; 
Kosenko et al.\ 2008; Yamaguchi et al.\ 2008a; Badenes et al.\ 2003; 2007). 
This can be explained by a stratified ejecta composition with Fe in the interior
that is theoretically expected (e.g., Gamezo et al.\ 2005; Maeda et al.\ 2010b) 
and has also been confirmed by optical observations of standard Type~Ia SNe 
(e.g., Tanaka et al.\ 2011). 
Although there are a few of well-established CC SNRs (e.g., Cas~A, W49B, and N132D) 
that exhibit relatively strong Fe K-shell emission, the Fe is highly ionized in all of them,  
unlike in G344.7--0.1 and other Type~Ia SNRs 
(Maeda et al.\ 2009; Ozawa et al.\ 2009; Xiao \& Chen 2008). 
Since highly-ionized atoms generally have higher emissivity than lowly-ionized ones, 
the large Fe flux in these CC SNRs is achieved even with relatively low abundance 
of this element.

\begin{table}[t]
\begin{center}
\caption{Masses of the hydrogen and $\alpha$-elements.
  \label{tab:mass}}
\begin{tabular}{lcccc}
\tableline\tableline
  Element  & \multicolumn{4}{c}{Mass (\Msun)}\\
  \cline{2-5}
  ~  &  Model Ax & Model Bx & Model As & Model Bs \\
  \tableline
  H & 164 & 175 & 153 & 166 \\
  Mg & 0.078 & 0.14 & 0.092 & 0.091 \\
  Si & 0.18 & 0.23 & 0.20 & 0.18 \\
  S  & 0.16 & 0.18 & 0.17 & 0.17 \\
  Ar & 0.043 & 0.048 & 0.046 & 0.048 \\ 
  Ca & 0.033 & 0.035 & 0.039 & 0.041 \\
  \tableline
\end{tabular}
\end{center}
\end{table}

\begin{figure*}[t]
  \begin{center}
	\includegraphics[height=4.5cm]{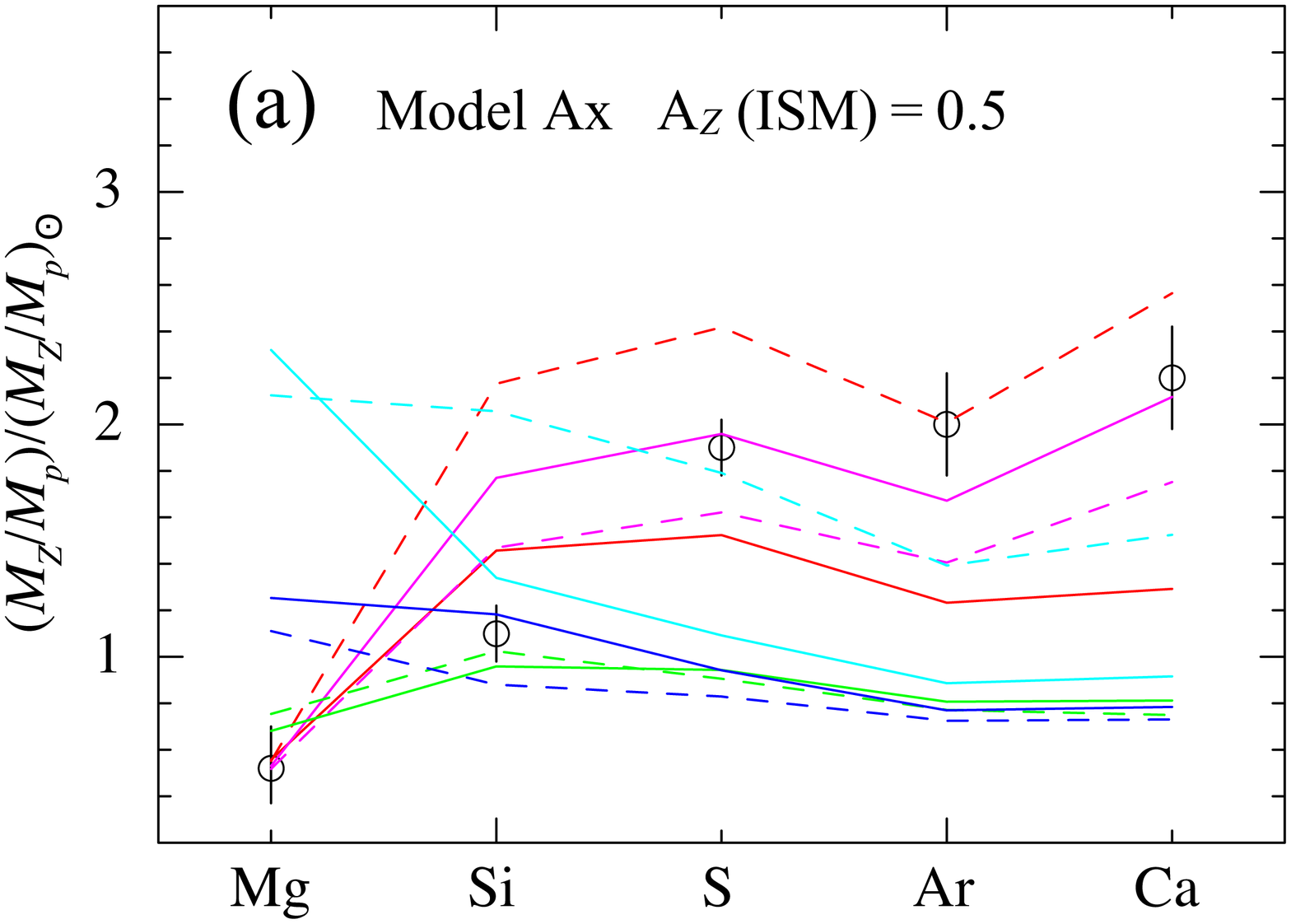}
	\includegraphics[height=4.5cm]{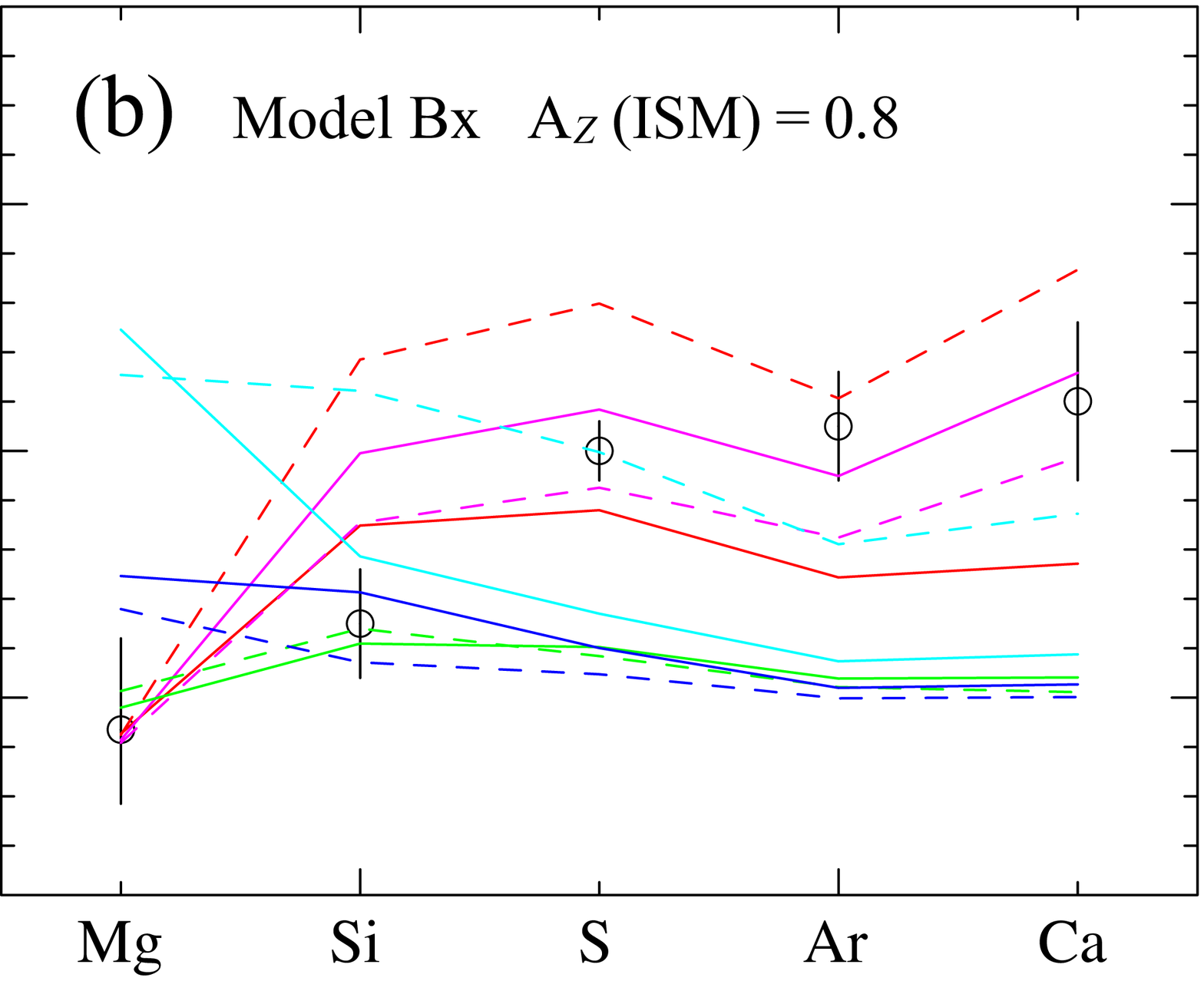}	
	\includegraphics[height=4.5cm]{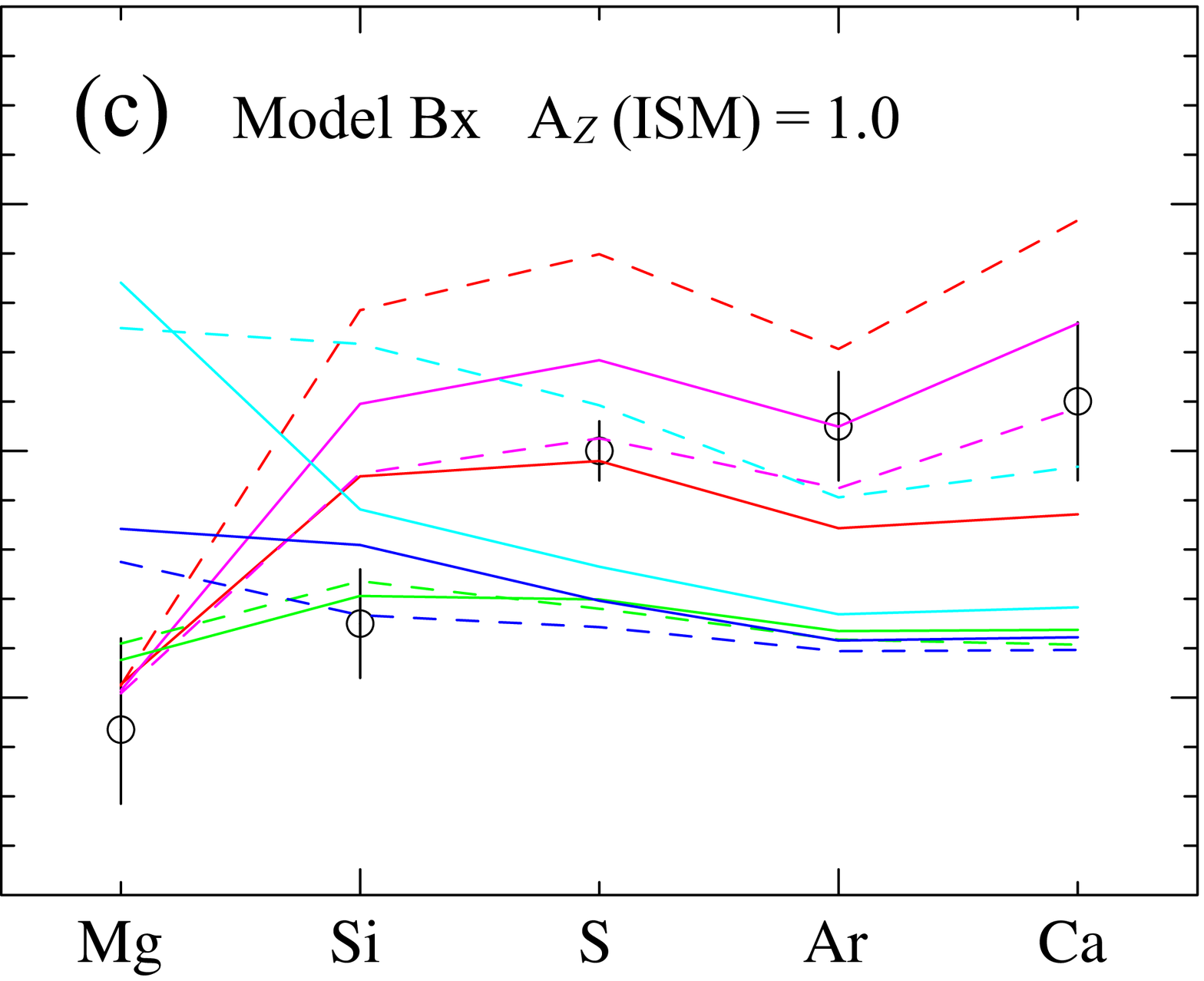}
\caption{Elemental mass abundances relative to the solar values compared with 
	those expected from several nucleosynthesis models for Type Ia SNe: 
	W7 (red solid lines), WDD1 (red dashed lines), WDD2 (magenta solid lines), 
	and WDD3 (magenta dashed lines) from Iwamoto et al.\ (1999), 
	and for CC SNe with different progenitor masses of 
	13\Msun\ (green solid lines), 15\Msun\ (green dashed lines), 
	18\Msun\ (blue solid lines), 20\Msun\ (blue dashed lines), 
	25\Msun\ (cyan solid lines), and 30\Msun\ (cyan dashed lines) 
	from Kobayashi et al.\ (2006). Note that contributions of the swept-up ISM 
	is taken into account (see text for details). The applied spectral models 
	and assumed ISM abundances are indicated in each panel. 
  \label{fig:mass}}
  \end{center}
\end{figure*}

The abundance pattern of the $\alpha$-elements we obtained is also 
supportive of the Type~Ia SN origin. 
The estimated masses of the hydrogen and $\alpha$-elements for each applied 
model are given in Table~\ref{tab:mass}, where a 3.$\!'$2-radius sphere for the X-ray 
emitting region, with a filling factor of unity, is assumed. 
Here, we compare these values with yields of different SN nucleosynthesis models. 
Since the large hydrogen mass of $\gtrsim$150\Msun\ implies that the SNR is 
dominated by the swept-up ISM, we take into account both the ejecta and ISM 
contributions. The total observed mass of the hydrogen or heavy elements $Z$ 
can be phenomenologically divided into those of the ejecta 
and ISM components as $M_{\rm H} = M_{\rm H, ej} + M_{\rm H, ISM}$ 
or $M_Z = M_{Z{\rm , ej}} + M_{Z{\rm , ISM}}$. 
($M_{\rm H, ej}$ should be 0 for Type Ia SNe.) 
The relation between $M_{Z{\rm , ISM}}$ and  $M_{\rm H, ISM}$ is given as 
\begin{equation}
  M_{Z{\rm , ISM}} = 
    A_Z{\rm (ISM)}~(m_Z/m_{\rm H})(n_Z/n_{\rm H})_{\odot}M_{\rm H, ISM},
\end{equation}
where $A_Z{\rm (ISM)}$ is the abundance of the ISM component (relative to the 
solar value) and the ratio $m_Z/m_{\rm H}$ corresponds to the average mass number 
of element $Z$. Therefore, we obtain 
\begin{equation}
\label{eq:mass}
  \frac{M_Z}{M_{\rm H}} = \frac{M_{Z{\rm , ej}}}{M_{\rm H}} + A_Z{\rm (ISM)}
    \left( \frac{m_Z}{m_{\rm H}} \right) \left( \frac{n_Z}{n_{\rm H}} \right)_{\odot}
    \left( 1- \frac{M_{\rm H, ej}}{M_{\rm H}} \right)
\end{equation}
for each element $Z$. 
In Figure~\ref{fig:mass}, we show the observed mass ratio $M_Z/M_{\rm H}$ of 
the different elements normalized by the solar values. 
On the other hand, theoretical values can be derived by substituting nucleosynthesis 
products of various SN models for $M_{\rm H, ej}$ and $M_{Z{\rm , ej}}$ in 
Equation~\ref{eq:mass}. 
Here we employ the Type Ia deflagration or delayed-detonation models 
(Iwamoto et al.\ 1999) and the CC SN models with various progenitor masses 
(Kobayashi et al.\ 2006). 
The representative results of the comparison are shown in Figure~\ref{fig:mass}, 
where for $A_Z{\rm (ISM)}$ we consider several values in the range of 0.5--1.0~solar. 
We find that the abundance pattern is broadly consistent with the Type~Ia models, 
but does not match with any CC models. 
We also considered several other nucleosynthesis models for CC SNe 
(e.g., Woosley \& Weaver 1995; Thielemann et al.\ 1996), and reached 
essentially the same conclusion. 
Although our estimates of the element masses is largely affected by an assumed 
distance and a filling factor, we note that the enhanced abundance ratios of 
S/Mg, Ar/Mg, and Ca/Mg can be explained only by an evolved Type~Ia SN. 
This is because no CC model predicts a Mg abundance less than a half of 
the heavier $\alpha$-elements' abundances.

A Type~Ia SNR in a dense environment (e.g., near molecular clouds) is quite 
rare (e.g., Andersen et al.\ 2011). 
In fact, \cite{g11} identified G344.7--0.1 as a CC SNR, based on its spatial association 
with HI clouds possibly created by recent ($\sim$$10^6$~yr ago) star formation. 
\cite{l11} was also supportive of its CC origin, since the strong asymmetry in 
its X-ray morphology was analogous to other CC SNRs. 
However, these preceding works lacked statistically sufficient spectral information, 
especially of Fe emission, that is so crucial in distinguishing SN types.
Similar situations have occurred with RCW~86 and N103B. 
RCW~86 was found to be evolving within a wind-blown bubble and thus suggested 
to be a CC SNR (Vink et al.\ 2006). Possible evidence of interaction with a molecular 
cloud was also pointed out in this SNR (Yamaguchi et al.\ 2008b). 
However, recent observations and a theoretical work strongly suggest its origin to be 
a Type~Ia SN (Williams et al.\ 2011; Yamaguchi et al.\ 2011b). 
N103B was also initially classified as a CC SNR because of its proximity to 
the young star cluster NGC~1850B with an age of $\sim$4~Myr 
(Chu \& Kennicutt 1988; Gilmozzi et al.\ 1994), 
but the {\it ASCA} and {\it Chandra} spectra revealed its Type~Ia SN origin 
(Hughes et al.\ 1995; Lewis et al.\ 2003). Moreover, Badenes et al.\ (2009) found 
that the progenitor was possibly associated with star formation at 50--150~Myr ago, 
suggesting no direct relationship between the SNR's progenitor and NGC~1850B. 
It is worth noting that Lewis et al.\ (2003) discovered that the EW maps of Si and S 
K-shell emission exhibit nearly symmetric shell-like structure, in contrast to 
the strong east--west asymmetry in the line emission measure (i.e., plasma density). 
In the morphological study by \cite{l11}, the symmetry of G344.7--0.1 was investigated 
only with the broad X-ray band image, which can be biased by density inhomogeneity 
in the ambient ISM. 
EW maps of specific elements could display an uncontaminated ejecta distribution, 
which could be more symmetric if G344.7--0.1 is indeed a Type~Ia remnant. 
This will be explored in our future work combining high-resolution images of 
{\it Chandra} and {\it XMM-Newton}.

The point-like soft X-ray source CXOU J170357.8--414302, located at the SNR's 
geometrical center, has been suggested to be either a CCO associated with the SNR 
or a foreground K0 star \citep{c10}. Since our result suggests a Type Ia origin of 
G344.7--0.1, the latter possibility is preferred.

From the morphological features in the radio and X-ray bands, G344.7--0.1 can be 
classified as a mixed-morphology (MM) SNR (Rho \& Petre 1998; see also \cite{g11} 
who revealed the complete shell structure in the radio including the faint eastern region). 
Despite continuous efforts by a number of authors (e.g., White \& Long 1991; 
Cox et al.\ 1999; Shelton et al.\ 1999; Petruk 2001; Slane et al.\ 2002), 
evolutionary characteristics which have led to the observed properties of 
MM SNRs are not well understood. 
It should be noted that all the MM SNRs for which progenitor types have been 
well identified are known to have CC origins. 
Therefore, if our conclusion about the progenitor type of G344.7--0.1 is correct, 
this result would demonstrate that Type~Ia SNe can also form remnants of 
this class\footnote{G337.2--0.7 is another possible MM SNR where a Type~Ia 
SN origin is suggested, but no evidence of Fe K-shell emission is reported 
(Rakowski et al.\ 2001; 2006).}$\!\!$. 
This may support an idea that the formation mechanism of this unusual morphology is 
not related to the nature of progenitors, but rather associated with their environment,  
as was assumed in the several previous works 
(e.g., White \& Long 1991; Shelton et al.\ 1999). 
Interestingly, recent {\it Suzaku} observations have discovered overionized plasmas 
in several MM SNRs (e.g., Yamaguchi et al.\ 2009; Ozawa et al.\ 2009), 
which are quite different from the ionization conditions in G344.7--0.1. 
Future works are warranted to comprehensively understand the evolution of 
both the morphology and ionization in MM SNRs.

The EW of the Fe-K$\alpha$ emission is not uniform but is enhanced in the southeast region 
(Section~\ref{ssec:image}). For more quantitative studies, we extracted spectra from 
two elliptical regions, ``SE'' and ``NW" shown in Figure~\ref{fig:image}(f), and fitted with 
the same models as used in Section~\ref{ssec:spectrum}. The Fe-K$\alpha$ EWs for 
SE and NW are 5.1 ($\pm 0.7$)~keV and 3.3 ($\pm 0.5$)~keV, respectively. 
If the temperature of the Fe ejecta is constant across the entire remnant, 
the surface brightness of the Fe emission is approximately proportional to 
$n_{\rm Fe}^2$. Therefore, the ejecta density in the SE region 
may be higher than that in the NW region by a factor of $\sim$1.2. 
An inhomogeneity of hot ejecta can arise from the interaction of SNRs with nonuniform 
ambient matter, as was observed, for example, in Kes~27 (Chen et al.\ 2008); 
the reflected shock after the collision between the blast wave and dense ISM can 
propagate backward and enhance the X-ray emission from the ejecta. 
In the case of G344.7--0.1, however, no evidence of dense matter is found 
around the SE region, but the NW rim is, in contrast, likely to have encountered 
a molecular cloud \citep{c10}. 
Therefore, one of the possible origins of  the enhancement is an inhomogeneous 
ejecta distribution caused by an asymmetric SN explosion. 
The small asymmetry in $^{56}$Ni production with a density variation of factor 
$\sim$1.2 is within the range that can be theoretically explained by an established 
delayed-detonation Type Ia SN model (e.g., Maeda et al.\ 2010a). 
Alternatively, an overall SE--NW temperature gradient can explain the inhomogeneity, 
but unfortunately, the electron temperature of the Fe ejecta component is difficult to 
constrain because of the relative weakness of its bremsstrahlung flux 
(see Figure~\ref{fig:full2}).

\subsection{Al Emission and ISM Metallicity}
\label{ssec:al}

The emission line of He-like Al at $\sim$1.6~keV has clearly been detected. This is 
the first detection of this element in the X-ray spectrum of an extended celestial source. 
Although we looked at spectra from several smaller regions of the SNR, no spatial 
concentration of this emission was found. We also carefully repeated the analysis 
with different background regions and confirmed that the spectral feature is not 
due to inaccurate background subtraction. 
Note that the intensity of neutral Al emission in the intrinsic background of the XIS is 
extremely low and stable compared to {\it XMM-Newton}/EPIC (Yamaguchi et al.\ 2006). 
We emphasize that this capability has enabled us to detect this emission robustly. 
{\it Chandra}/ACIS has a similarly low background level, but the effective area 
around this energy is about a factor of two smaller than that of the XIS. 
The energy resolution is also lower in the ACIS, which makes 
detection of weak emission more difficult.

Since Al emission line data are not available in the {\tt vnei} code in {\tt XSPEC}, 
we refer to AtomDB\footnote{http://www.atomdb.org/} and calculate the emissivity 
for the He-like Al-K blend to estimate the Al abundance, $A_{\rm Al}$. 
The relation between the line flux $F_{\rm He\mathchar`-like}$ and 
emissivity $\varepsilon _{kT_e, n_et}$ for He-like Al is given as
\begin{equation}
 F_{\rm He\mathchar`-like} = \varepsilon _{kT_e, n_et}~A_{\rm Al}~(n_{\rm Al}/n_{\rm H})_{\odot}{\rm VEM}, 
\end{equation}
where $(n_{\rm Al}/n_{\rm H})_{\odot}$ is the number ratio of Al to H in the solar 
(Anders \& Grevesse 1989). 
For the NEI plasma with the best-fit $kT_e$ and $n_et$ values for Models~Ax and Bx 
(Table~\ref{tab:best}), we obtain $\varepsilon$ = $1.4 \times 10^{-11}$~cm$^3$~s$^{-1}$ 
and $1.3 \times 10^{-11}$~cm$^3$~s$^{-1}$, respectively. 
Therefore, the flux of the 1.6-keV Gaussian corresponds to an Al abundance of 
$1.3_{-0.3}^{+0.4}$~solar for Model~Ax and $1.8_{-0.6}^{+0.8}$~solar for Model~Bx, 
slightly higher than the solar value. 
Since a Type~Ia SN scarcely produces Al (Iwamoto et al.\ 1999), this abundance 
enhancement indicates either (1) the ambient ISM has been enriched with Al by 
preceding CC SNe, or (2) the SNR really has a CC origin against our argument 
in Section~\ref{ssec:fe}. We consider the former is the case as we discuss below.

It has been suggested that chemical yields of $^{27}$Al and other 
neutron-rich elements in CC SNe strongly depend on the metallicity 
of progenitor stars (e.g., Kobayashi et al.\ 2006). 
This is because these elements are enhanced upon SN explosive nucleosynthesis 
by excess neutrons in $^{22}$Ne, which is transformed from $^{14}$N during 
the He-burning phase of the progenitor (see Chapter~9 of Arnett 1996). 
Since both Mg and Al arise from C/Ne burning, the Al-to-Mg abundance ratio 
can be used to indicate the metallicity. In fact, a strong metallicity dependence 
of the Al/Mg ratio has been confirmed by several stellar observations 
(Gehren et al.\ 2004; 2006; Andrievsky et al.\ 2010 and references therein). 
Notably, most stars with a sub-solar Mg abundance (Mg/H = 0.5--1~solar) exhibit 
enhanced Al/Mg values of 1--2~solar  (see Figure~9 of Andrievsky et al.\ 
2010\footnote{Although Andrievsky et al.\ (2010) refer to Lodders (2003) for 
solar abundance values, the Al/Mg ratios in Anders \& Grevesse (1989) 
and Lodders (2003) are almost identical.}), suggesting that  the Al abundance 
in the Sun is anomalously low (and/or the Mg abundance is high) 
compared to Galactic nearby stars. 
In metal-poor stars (Mg/H $<$0.1~solar), on the other hand, 
the Al/Mg ratio is depleted to less than 0.3~solar.

\begin{figure}[t]
  \begin{center}
      \includegraphics[scale=.345]{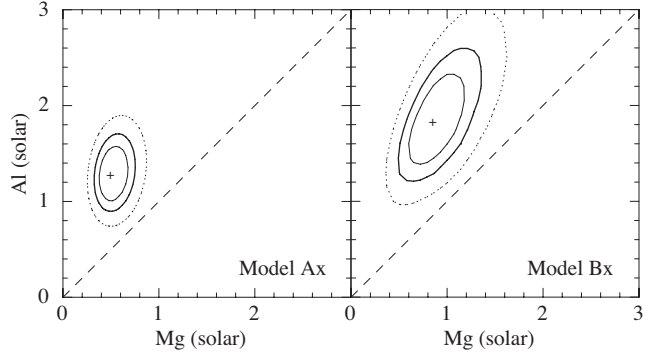}
\caption{Confidence contours at 68\% (thin lines), 90\% (thick lines), and 
	99\% (thin dotted lines) for the abundance ratio of Al to Mg. 
	The solar abundance ratio is shown by the dashed lines.
  \label{fig:cont}}
  \end{center}
\end{figure}

Figure~\ref{fig:cont} shows confidence contours for the abundance of Al relative to 
that of Mg. We obtain the Al/Mg abundance ratios of $2.5_{-1.0}^{+1.5}$~solar and 
$2.1_{-0.7}^{+1.1}$~solar for Models~Ax and Bx, respectively. 
The ratios are, furthermore, found to be higher than unity at $>$99\% confidence 
in both cases and even more significant in the case of Models~As and Bs. 
To estimate systematic uncertainties, we fitted the spectra with the same plasma 
models but replaced the {\tt wabs} component with an updated absorption model of 
Wilms et al.\ (2000; {\tt tbabs}) where photon extinction by interstellar dust grains is 
additionally taken into account. We found that this change did not affect significantly 
the measurement of the Al/Mg ratio. 
We also repeated the spectral fitting by assuming different Fe abundances 
in the ``NEI plasma'' component (which can be regarded as an Fe abundance in 
the ISM), and obtained no significant difference in the result. 
The enhanced relative abundance of Al/Mg as well as the slightly-depleted value 
of the absolute Mg abundance are consistent with the chemical characteristics in 
the typical Galactic stars. 
Therefore, we propose that the Mg and Al emission from G344.1--0.7 originates 
predominantly from the blast-shocked ISM with the metallicity of a solar-like value 
(not less than 0.1~solar). The moderate metallicity of the ambient ISM can be 
due to preceding CC SNe in nearby regions. 
However, there is still a large difficulty in the abundance measurement 
of Al due to the limited energy resolution (see Section~\ref{ssec:future}), 
and thus future efforts are essential for more definitive conclusion.

\subsection{Cr and Mn Emissions and Progenitor's Metallicity}
\label{ssec:crmn}

We have also found possible evidence of Cr and Mn emission. The centroids of these 
lines are fully consistent with those observed in Tycho's SNR (Tamagawa et al.\ 2009). 
We should note, however, that there are many sources of systematic uncertainties 
we have not yet taken into account. 
For instance, we simply assume a single-temperature plasma for the NEI plasma 
component. The value of $kT_e$ is, therefore, determined mainly by the 
K$\beta$/K$\alpha$ intensity ratios of Si and S, since these lines have 
a large number of photon statistics. 
If we allow multiple temperatures, which can be the case for most evolved SNRs, 
the continuum level increases slightly around the Cr and Mn features. 
This results in slightly lower fluxes for the weak line components; 
for example, adding another NEI plasma with a different electron temperature to 
Model~Bx or Bs, we obtained a $\sim$20\% lower flux of the Cr-K$\alpha$ emission. 
Therefore, the detection of these features is still marginal, and we absolutely need 
deeper and higher energy-resolution observations to confirm their presence.

Nevertheless, it would be meaningful to make a rough estimate of the mass ratio of 
Mn to Cr, because this ratio is suggested to be a good metallicity tracer for Type~Ia 
progenitors (Badenes et al.\ 2008) for reasons similar to those mentioned above. 
The progenitor's metallicity should not necessarily be consistent with that of 
the swept-up ISM (i.e., the source of Mg and Al), since the ISM metallicity is likely to 
have been modified by recent CC SNe after the Type~Ia progenitor had been created. 
Using the best-fit values of Model~Bx (Table~\ref{tab:best}), we derive the Mn/Cr flux 
ratio to be $0.67_{-0.48}^{+0.55}$. Given that the ionization states of both the elements 
are lower than He-like, similar to Fe, the Cr and Mn ions are likely to be associated with 
the Fe-rich plasma. Thus, the electron temperature and ionization timescale should be 
in the ranges of 2--10~keV and (6--12)\,$\times$\,$10^9$~cm$^{-3}$~s, respectively. 
According to Figure~4 of Badenes et al.\ (2008), the emissivity ratio of 
$\varepsilon _{\rm Mn}$/$\varepsilon _{\rm Cr}$ is estimated to be 0.5--0.75. 
Therefore, the Mn-to-Cr mass ratio is constrained in the range of 0.1--1.0. 
Since the relation between the mass ratio and progenitor's metallicity is given as 
$M_{\rm Mn}$/$M_{\rm Cr}$ = $5.3 \times Z^{0.65}$ (Badenes et al.\ 2008), 
we obtain $Z$ = 0.002--0.07. Although the uncertainty is relatively large, 
this does not conflict with the single-degenarate scenario as a candidate for 
the Type~Ia progenitor system, where the metallicity must be 
higher than $Z \sim 0.001$ for the ``optically thick wind'' to be driven effectively 
(Kato \& Hachisu 1994; Hachisu et al.\ 1996; Kobayashi et al.\ 1998).

\subsection{Future Prospect}
\label{ssec:future}

\begin{figure}[t]
  \begin{center}
      \includegraphics[scale=.36]{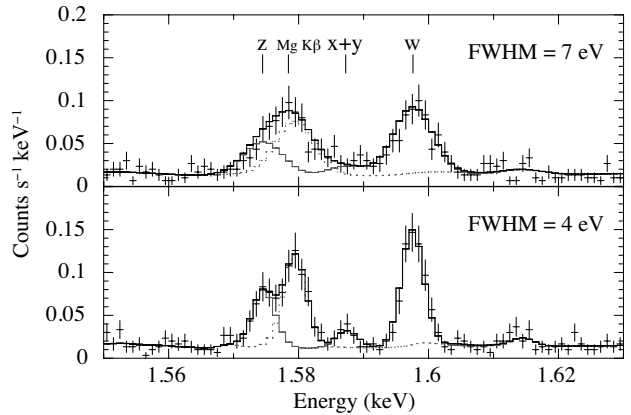}
\caption{Simulated {\it ASTRO-H}/SXS spectra for G344.7--0.1 with the total 
	exposure of 300~ks, in the energies around the Al-K$\alpha$ lines. 
	The FWHM of 7~eV (top) or 4~eV (bottom) is assumed. 
	Resonance (w), intercombination (x+y), and forbidden (z) lines of Al, 
	and He-like Mg-K$\beta$ line are clearly separated in the bottom.
	  \label{fig:sxs}}
  \end{center}
\end{figure}

Finally, we briefly mention the remaining problems and prospect for future observations. 
Although we have detected the possible feature of Al emission, our abundance estimates 
is still inconclusive. With the energy resolution of semiconductor detectors (e.g., the XIS), 
the He-like Al-K$\alpha$ line is totally blended with the Mg-K$\beta$ line, and 
both of them as well as the Mg-K$\alpha$ line are blended with Fe L-shell emissions. 
Since the L-shell emissivities are generally uncertain due to the limitation of atomic data, 
determination of the Mg and Al abundances can be systematically affected. 
Interstellar dust scattering of X-ray photons out of the line of sight (e.g., Draine 2003) 
and abundance inhomogeneity in the foreground ISM, which were not taken into 
account in this paper, are potential additional sources of the uncertainties. 
Moreover, nonuniform temperatures and ionization timescales, expected in evolved 
SNRs such as G344.7--0.1, also complicate the abundance measurement.

In 2014, {\it ASTRO-H} (Takahashi et al.\ 2010) will be launched and will carry 
a micro-calorimeter (Soft X-ray Spectrometer: SXS) with greater spectral resolution 
for diffuse objects than has been available with previous instruments. 
This will enable us to resolve the K-shell lines of Mg and Al from the Fe L-shell lines 
and to measure the abundances much more accurately. 
In Figure~\ref{fig:sxs}, we show examples of simulated spectra for the brightest 
part of G344.7--0.1. The full width at half maxima (FWHM) are assumed to be 
7~eV and 4~eV, which are the ``requirement" and ``goal" designs for the SXS, 
respectively (Mitsuda et al.\ 2010). 
In the case that the latter resolution is achieved, the fluxes of the Al-K$\alpha$ and 
Mg-K$\beta$ lines can be determined almost independently. 
We can also expect that emission from low-abundance elements, like Al, will be 
observed in a large number of other SNRs. Diagnostics using such weak lines 
will provide us fruitful information about detailed nature of SNRs' progenitor, 
environment, and SN explosion mechanism.

\acknowledgments

The authors appreciate a number of constructive suggestions from the anonymous referee.
We are also grateful to Drs.\ Laura A.\ Lopez and Kazimierz J.\ Borkowski for helpful 
discussions and Dr.\ Yoh Takei for providing information about {\it ASTRO-H}/SXS.
H.Y.\ is supported by Japan Society for the Promotion of Science (JSPS) Research Fellowship 
for Research Abroad, and acknowledges funding from NASA ADP grant NNX09AC71G.

\bigskip


\end{document}